\documentclass[conference,a4paper]{IEEEtran}
\usepackage{amsmath,amssymb,amsfonts}
\usepackage{graphicx}
\usepackage{float}
\usepackage{subfig}
\usepackage{color}
\usepackage{cite}
\usepackage{multirow}
\usepackage{algorithm}  
\usepackage{algorithmicx}  
\usepackage{algpseudocode} 
\usepackage{bm}
\usepackage[colorlinks,linkcolor=black,anchorcolor=black,
citecolor=black,urlcolor=black]{hyperref}

\usepackage{stfloats}

\usepackage{siunitx}
\DeclareSIUnit\bps{bps}
\DeclareSIUnit\Torr{Torr}
\DeclareSIUnit\torr{Torr}
\DeclareSIUnit\sample{Sa}
\newcommand*{\circled}[1]{\lower.7ex\hbox{\tikz\draw (0pt, 0pt)%
  circle (.5em) node {\makebox[1em][c]{\small #1}};}}

\ifCLASSINFOpdf

\else

\fi

\graphicspath{{figures/}}

\begin{document}

\title{300 GHz Channel Measurement and Characterization in the Atrium of a Building}

\author{
\IEEEauthorblockN{Yuanbo~Li\IEEEauthorrefmark{1}, Yiqin Wang\IEEEauthorrefmark{1}, Yi Chen\IEEEauthorrefmark{2}, Ziming Yu\IEEEauthorrefmark{2}, and Chong~Han\IEEEauthorrefmark{1}}
\IEEEauthorblockA{\IEEEauthorrefmark{1} Terahertz Wireless Communications (TWC) Laboratory, Shanghai Jiao Tong University, China \\ Email:  \{yuanbo.li,wangyiqin,chong.han\}@sjtu.edu.cn\\
\IEEEauthorrefmark{2} Huawei Technologies Co., Ltd, China.
Email: \{chenyi171,yuziming\}@huawei.com
}
}

\maketitle

\begin{abstract}
With abundant bandwidth resource, the Terahertz band (0.1~THz to 10~THz) is envisioned as a key technology to realize ultra-high data rates in the 6G and beyond mobile communication systems. However, moving to the THz band, existing channel models dedicated for microwave or millimeter-wave bands are ineffective. To fill this research gap, extensive channel measurement campaigns and characterizations are necessary. In this paper, using a frequency-domain Vector Network Analyzer (VNA)-based sounder, a measurement campaign is conducted in the outdoor atrium of a building in 306-321~GHz band. The measured data are further processed to obtain the channel transfer functions (CTFs), parameters of multipath components (MPCs), as well as clustering results. Based on the MPC parameters, the channel characteristics, such as path loss, shadow fading, K-factor, etc., are calculated and analyzed. The extracted channel characteristics and numerology are helpful to study channel modeling and guide system design for THz communications.
\end{abstract}

\IEEEpeerreviewmaketitle

\section{Introduction}
\par To realize the promising yet thrilling applications, such as metaverse, digital twin, etc., data rates in the sixth generation (6G) and beyond mobile communication systems are expected to exceed hundreds of gigabits per second and even Terabits per second~\cite{chen2021terahertz}. To sustain the explosively grown data traffic, the Terahertz (THz) band, ranging from \SI{0.1}{THz} to \SI{10}{THz} that offers abundant bandwidth resource, is envisioned as a key technology~\cite{akyildiz2018combating,rappaport2019wireless}. 
However, to realize THz communications, one major challenge lies on the channel modeling in the THz band, which requires extensive measurement campaigns to extract the channel characteristics. With developments of THz hardware, recently a few research groups have built up THz channel sounders and conducted channel measurement campaigns. 

The THz channel study can be {dated back to 2011, when the research group in Technische Universität Braunschweig reported many measurement results in both indoor scenarios like conference room, office, etc., and outdoor scenarios such as vehicle-to-vehicle channels~\cite{priebe2011channel,priebe2013ultra,Eckhardt2021channel}.} Recently, the research group in New York University has made many progresses in channel characterizations in \SIrange{140}{142}{GHz} band in various scenarios, such as office, factory, urban, etc.~\cite{ju2021millimeter,ju2022sub,ju2021sub}. Moreover, the research group from University of Southern California has measured THz channels in frequency bands near \SI{140}{GHz} and \SI{220}{GHz}, mainly in outdoor urban scenarios~\cite{abbasi2020double,abbasi2021double}. Based on VNA-based channel sounder, we have conducted several channel measurement campaigns and extracted channel characterizations in frequency bands near \SI{140}{GHz} and \SI{300}{GHz}, in indoor scenarios like conference room, hallway, and corridor in Shanghai Jiao Tong University (SJTU)~\cite{chen2021channel,wang2022thz,li2022channel}. However, according to the authors' best knowledge, there is still no channel measurement campaign investigating outdoor channels at \SI{300}{GHz}, where we devote to fill this research gap.
\par In this paper, a measurement campaign is conducted in the atrium of a building on the SJTU campus, focusing in \SIrange{306}{321}{GHz} band by using a frequency-domain VNA-based channel sounder. Overall, 21 Rx positions are measured with {over 3700 channel transfer functions (CTFs)} to fully characterize the THz wave propagation in the atrium. Based on the measured data, post-processing procedures, including calibration, channel estimation, and MPC clustering, are further implemented to obtain CTF, MPC parameters, as well as clustering results. Finally, the channel characteristics, including the path loss, shadow fading, K-factor, delay spread, angular spreads, as well as cluster parameters, are investigated in depth with a complete characteristic table. The measured results not only reveal the strong sparsity of THz channels, but also establish numerology useful for channel modeling and system designs in the THz band.
\par The remainder of the paper is organized as follows. In Sec.~\ref{sec:measurement}, the VNA-based channel sounder, measurement set-up and measurement deployment are explained in detail. Furthermore, the data processing procedure is introduced in Sec.~\ref{sec:clustering}. In light of the measurement results, the channel characteristics are analyzed in Sec.~\ref{sec:char}. Finally, Sec.~\ref{sec:conclude} concludes the paper. 

\section{Channel Measurement Campaign}
\label{sec:measurement}
\par  In this section, the measurement campaign in the atrium scenario is described, including the measurement system, the measurement set-up and the measurement deployment.
\subsection{Measurement System}
Our measurement system is based on a VNA-based channel sounder, whose measurement ability covers any frequency bands within \SIrange{260}{400}{GHz}. Moreover, both transmitter (Tx) and receiver (Rx) modules are installed on rotators, lifters, as well as electrical carts, which makes it easy for us to change the steering angles, heights and locations of Tx and Rx modules. For full-fledged description of our measurement system, readers are encouraged to refer to~\cite{yuanbo2022channel}.
\subsection{Measurement Setup}
\begin{table}[tbp]
	\centering  % 显示位置为中间
	\caption{Measurement parameters}  % 表格标题
	\label{tab:parameters}  % 用于索引表格的标签
	%字母的个数对应列数，|代表分割线
	% l代表左对齐，c代表居中，r代表右对齐
	\begin{tabular}{l l}  
		\hline  % 表格的横线
		& \\[-6pt]  %可以避免文字偏上来调整文字与上边界的距离
		Parameter & Values\\
		\hline
		& \\[-6pt]  %可以避免文字偏上 
		Frequency band & 306-\SI{321}{GHz}\\

		Bandwidth & \SI{15}{GHz}\\

		Sweeping interval& \SI{2.5}{MHz}\\

		Sweeping points & 6001\\

		Maximum delay& \SI{400}{ns}\\

		Maximum path length& \SI{120}{m}\\

		Time resolution& \SI{66.7}{ps}\\

		Space resolution& \SI{2}{cm}\\

        Tx height&\SI{2.2}{m}\\
        
        Rx height&\SI{1.5}{m}\\
        
        Antenna gain of Tx & \SI{7}{dBi}\\
        
		Antenna gain of Rx & \SI{25}{dBi}\\
		
		HPBW of Tx antenna & $30^\circ$\\

		HPBW of Rx antenna & $8^\circ$\\

		Rx azimuth rotation range&[$0^\circ:10^\circ:360^\circ$]\\

		Rx elevation rotation range&[$-20^\circ:10^\circ:20^\circ$]\\

		\hline
	\end{tabular}
	\vspace{-0.5cm}
\end{table}
\par The measurement setups are summarized in Table.~\ref{tab:parameters}, which are introduced in detail as follows. The frequency band that we measure ranges from \SI{306}{GHz} to \SI{321}{GHz}, across a \SI{15}{GHz} bandwidth. Moreover, the sweeping interval in the frequency domain is \SI{2.5}{MHz}, corresponding to a \SI{400}{ns} maximum delay and \SI{120}{m} maximum path length of MPCs. Furthermore, the measurement system has a \SI{66.7}{ps} time resolution, which suggests that MPCs whose path length difference is larger than \SI{2}{cm} can be separated. The heights of Tx and Rx are \SI{2.2}{m} and \SI{1.5}{m}, respectively. 

Besides, to obtain a large beamwidth for wide coverage, the transmitter is only equipped with a standard waveguide WR2.8, which has \SI{7}{dBi} antenna gain and a $30^\circ$ half-power beamwidth (HPBW). By contrast, the Rx is equipped with a directional antenna with a \SI{25}{dBi} antenna gain and a $8^\circ$ HPBW. To capture MPCs from various directions, the Rx scans the spatial domain with $10^\circ$ angle steps, from $0^\circ$ to $360^\circ$ in the azimuth plane and $-20^\circ$ to $20^\circ$ in the elevation plane. {Therefore, there are totally 180 CTFs measured at each Rx position with different steering angles of Rx.}
\subsection{Measurement Deployment}
\begin{figure}
    \centering
    \subfloat[Pictures of the atrium.] {
     \label{fig:corridor}     
    \includegraphics[width=0.8\columnwidth]{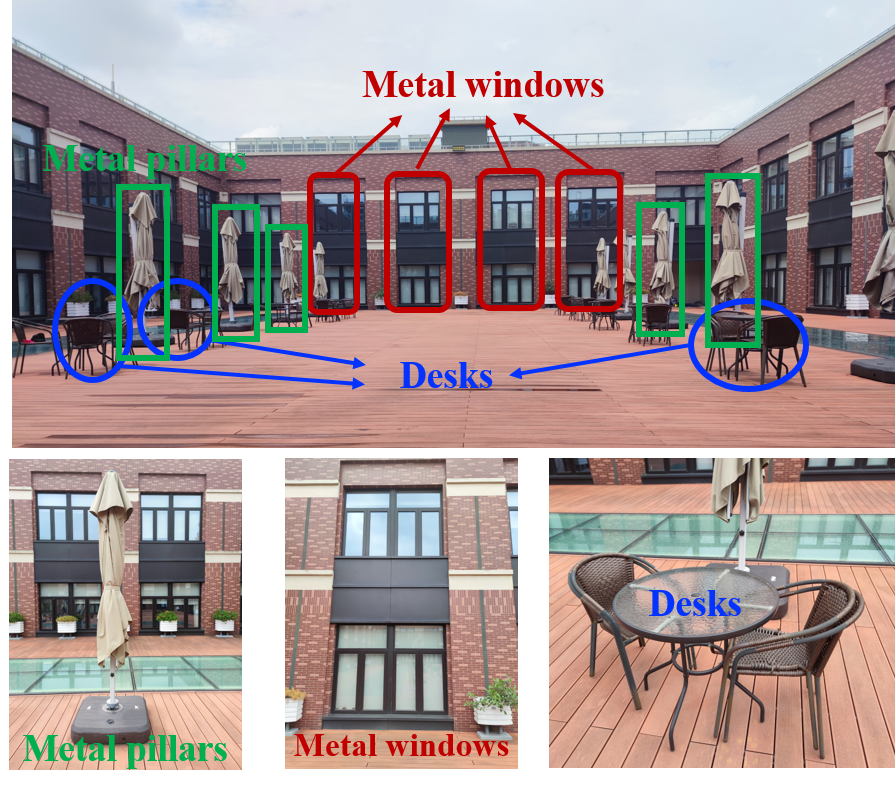}  
    }
    \quad
    \subfloat[Panoramic photo.] {
     \label{fig:panoramic}     
    \includegraphics[width=0.8\columnwidth]{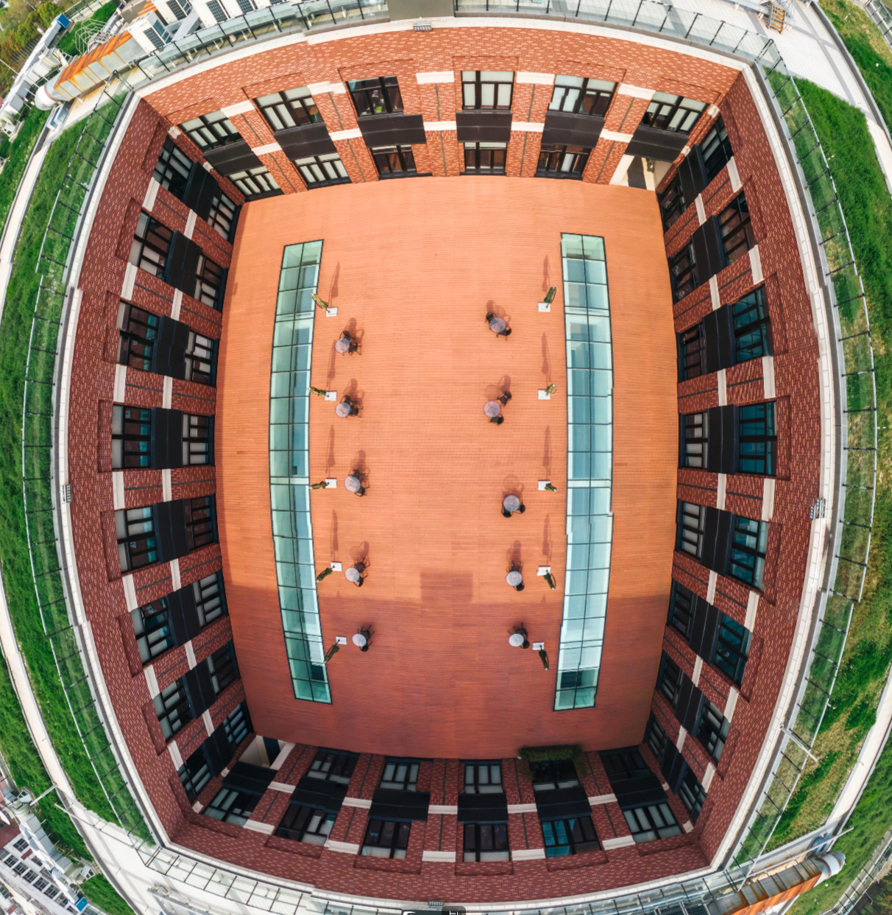}  
    }
    \caption{Pictures of the atrium scenario.}
    \label{fig:pic}
    \vspace{-0.5cm}
\end{figure}
\begin{figure}
    \centering
    \includegraphics[width=0.7\columnwidth]{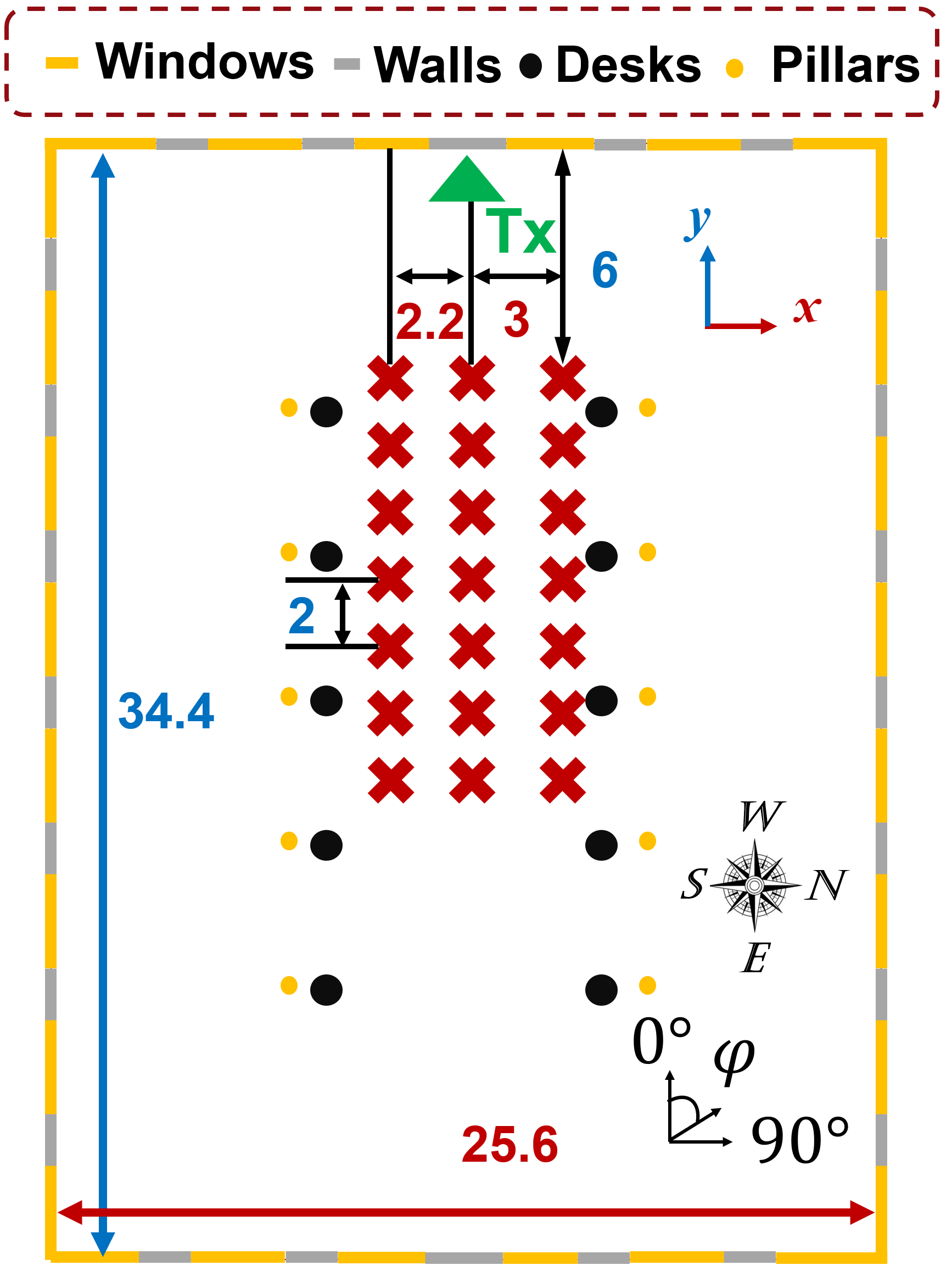}  
    \caption{The layout of the measurement campaign conducted in the atrium of a building.}
    \label{fig:layout}
    \vspace{-0.5cm}
\end{figure}
\par The measurement campaign is conducted in an outdoor atrium on the fourth floor of the Longbin Building in Shanghai Jiao Tong University, as shown in Fig.~\ref{fig:pic}. In the center area of the atrium, there are many glass desks and metal pillars. Moreover, on the outskirts of the atrium, there are four concrete walls, furnished with glass windows and metal window frames. The floor of the atrium is mostly made with wood, except two small areas furnished with glass. Furthermore, The length and width of the atrium are 34.4 m and 25.6 m, respectively.
\par The transmitter and receiver are located in the center area of the atrium, as shown in Fig.~\ref{fig:layout}. The transmitter remains static, deployed near the middle of the western wall of the atrium, pointing to the eastern direction. Besides, 21 receiver positions are selected, separated into three columns. Two consecutive receiver positions in each column are separated with a 2 m interval. As a result, the range of the separation distance between Tx and Rx is approximately \SIrange{6}{18}{m}. Last but not least, all Rx positions have LoS propagation. {For each Rx point, it takes around 45 minutes to measure the channel. Overall, there are 3780 CTFs that are measured.}
\section{Data Processing Procedure}
\label{sec:clustering}
In this section, data post-processing procedures are introduced, including the calibration, channel estimation, and MPC clustering. The measured $S_{21}$ parameters from VNA include not only the CTFs of THz channel, but also noises and influences from the measurement system, which are eliminated through the calibration process. Furthermore, after obtained the CTFs of the THz channel, the modified space-alternating generalized expectation-maximization (SAGE) algorithm is used to accurately estimate the parameters of MPCs~\cite{yuanbo2022channel}. Moreover, the Density-Based Spatial Clustering of Applications with Noise (DBSCAN) algorithm is selected to cluster the extracted MPCs~\cite{ester1996density}. 
\label{sec:dataprocess}
\subsection{Calibration}
\label{sec:calib}
To eliminate influences of measurement systems, two measurements of S parameter are conducted with different setups. The first one is the real measurement, i.e., we measure the S parameters when Tx/Rx are located in certain positions, which includes the effects of the THz channel as well as the unwanted effects of cables and other things. The second one is to measure the S parameters when Tx/Rx modules are directly connected, which only reflects the unwanted factors. As a result, the CTF of the THz channel can be expressed as,
\begin{equation}
    H=\frac{S_{21}^{\text{measure}}}{S_{21}^{\text{extra}}S_{21}^{\text{connect}}}
\end{equation}
where $S_{21}^{\text{extra}}$ represents influences of components due to the different set-ups when measuring $S_{21}^{\text{measure}}$ and $S_{21}^{\text{connect}}$. For example, horn antennas are used when measuring $S_{21}^{\text{measure}}$ and not used for $S_{21}^{\text{connect}}$. 
\subsection{Channel Estimation}
\par The measured CTFs of the THz channel are superposition of many MPCs, as
\begin{equation}
\begin{split}
    \bm{H}=\sum_{l=1}^L\bm{s}_l(k)+\bm{W}
    \label{eq:cd}
\end{split}
\end{equation}
where $\bm{H}$ denotes the $N_r\times K$ CTF matrix, with $N_r$ denoting number of scanning directions of Rx and $K$ representing number of samples in the frequency domain. Moreover, $\bm{W}$ stands for the noise term. For the $l^\text{th}$ MPC, its influence $\bm{s}_l(k)$ is expressed as
\begin{equation}
    \bm{s}_l(k)=\alpha_l\bm{c}_r(\bm{\Omega}_{r,l})^\text{T}\text{e}^{j\bm{\phi}_{l}}\text{e}^{-j2\pi f_k\tau_l}
\end{equation}
where $\alpha_l,\tau_l,\bm{\Omega}_{r,l}$ denote the real-valued path gain, time-of-arrival (ToA), and direction-of-arrival (DoA) of the $l^\text{th}$ MPC, respectively. $\bm{\phi}_{l}$ represents the phase of the $l^\text{th}$ component due to effects of antennas, reflections, etc., which is related to the steering angle of Rx. $f_k$ stands for the carrier frequency at the $k^\text{th}$ sampling point. $\bm{c}_{r,n_r}(\cdot)$ stands for the real-valued radiation pattern of the Rx antenna. The direction $\bm{\Omega}_{r,l}$ is dependent on the azimuth angle-of-arrival (AoA) and elevation angle-of-arrival (EoA), as
\begin{equation}
    \bm{\Omega}_{r,l}=[\cos(\varphi_{r,l})\cos(\theta_{r,l}),\sin(\varphi_{r,l})\cos(\theta_{r,l}),\sin(\theta_{r,l})]^\text{T}
\end{equation}
where $\varphi_{r,l}$ and $\theta_{r,l}$ are AoA and EoA of the $l^\text{th}$ MPC, respectively.
\par Based on the measured CTF matrix in~\eqref{eq:cd}, the modified SAGE algorithm in~\cite{yuanbo2022channel} is utilized to extract the parameters of MPCs, including their ToA, AoA, EoA and power. Specifically, only MPCs with power larger than a threshold $P_{\text{th}}$ is estimated, which is set as
\begin{equation}
    P_\text{th} [\text{dB}]=20\log_{10}\alpha_1-30
\end{equation}
where $\alpha_1$ is the path gain of the strongest path at a certain Rx position.
\subsection{MPC Clustering}
\label{sec:mpcc}
\par In realistic channels, clusters emerge since many MPCs appear with similar parameters. For instance, a reflected MPC usually occurs with many surrounding scattering MPCs that have similar ToAs and DoAs. To study the cluster behaviour of the THz channel, the DBSCAN algorithm is selected for MPC clustering~\cite{ester1996density}. In particular, it is found that the clustering performance is better by adopting the multi-path component distance (MCD) as the distance metric instead of Euclidean distance, which is defined as~\cite{chen2021channel}
\begin{equation}
\begin{split}
    \text{MCD}_{i,j}=\sqrt{\xi\frac{(\tau_i-\tau_j)^2}{\tau_m^2}+||\bm{\Omega}_{t,i}-\bm{\Omega}_{t,j}||^2_2}
\end{split}
\end{equation}
where $\xi$ denotes a weighting factor that controls the weight of the ToA in the MCD, which is set as 3 in this work.
\par The parameters of clusters, namely their ToA, AoA, EoA and power, are determined in the following manner. The ToA, AoA and EoA of the clusters as those of the strongest path in the cluster, while cluster power is the sum of the power of MPCs within the cluster.
\section{Channel Characterization and Analysis}
\label{sec:char}
\par In this section, the channel characteristics, including the path loss, shadow fading, K-factor, delay spread, angular spreads, as well as cluster parameters, are calculated and analyzed. 
\subsection{Path Loss and Shadow Fading}
\begin{table*}[tbp]
    \centering
    \caption{Channel characteristics in the corridor scenario in the THz band.}
    \includegraphics[width = 1.85\columnwidth]{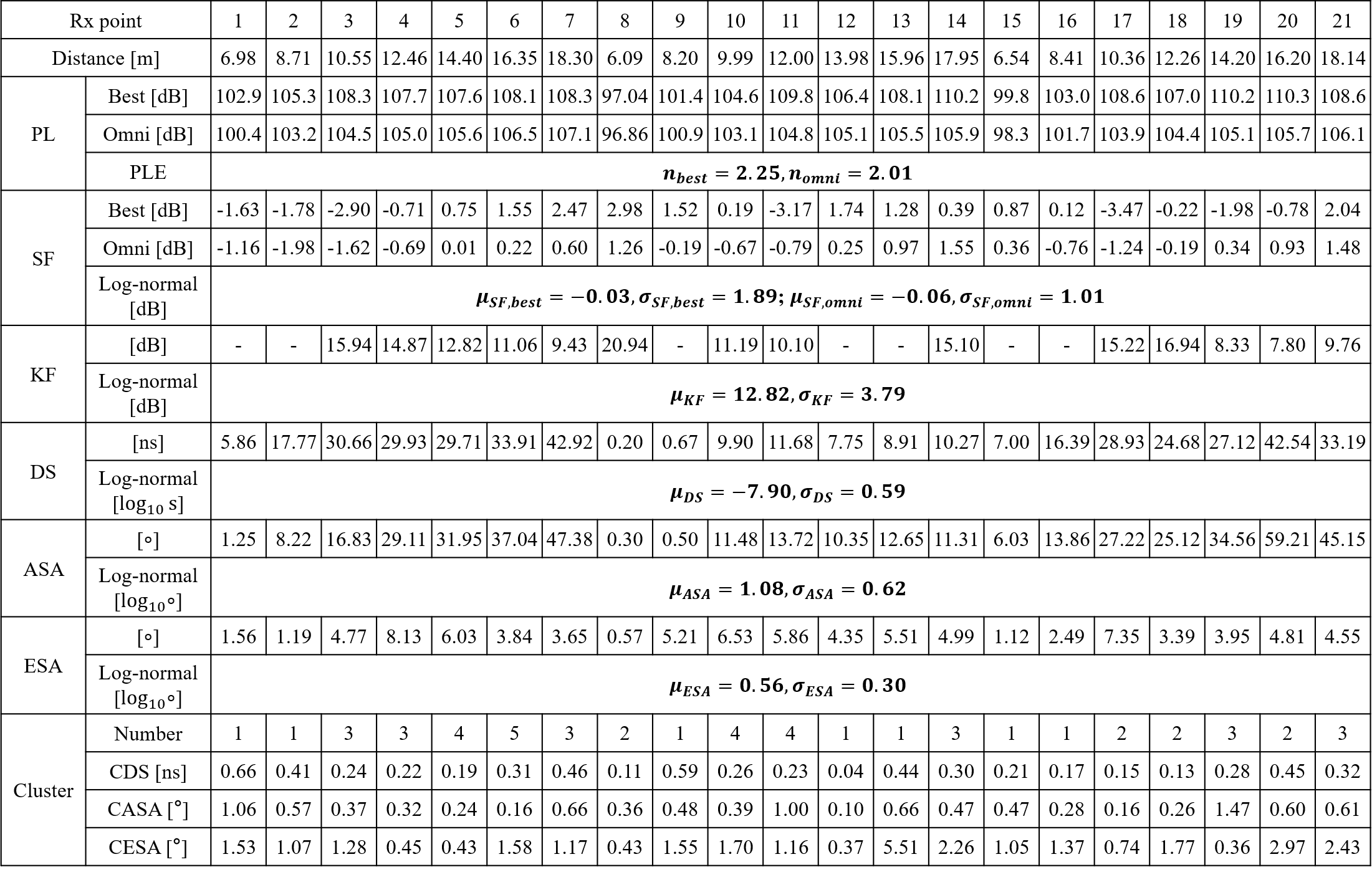}
    \label{tab:char}
    \vspace{-0.5cm}
\end{table*}
\par The path loss of the THz channel is normally evaluated in terms of the best direction path loss and the omnidirectional path loss. The best direction path loss represents the path loss on the pointing direction of Rx antennas with the strongest received power, while the omnidirectional path loss is related to the overall received power from all MPCs, respectively as
\begin{align}
    \text{PL}_{\text{best}} ~[\text{dB}]&=-10\log_{10}(\max_{n_r}{\frac{1}{K}\sum_{k=1}^K\left|H_{n_r}[k]\right|^2})\\
    \text{PL}_{\text{omni}} ~[\text{dB}]&=-10\log_{10}(\sum_{l=1}^L{\left|\alpha_{l}\right|^2})
\end{align}
where $H_{n_r}$ is the CTF for $(n_r)^\text{th}$ scanning direction at Rx.
\par Generally speaking, path loss follows a linear function with respect to the Euclidean distance between Tx and Rx, which can be modeled by using a close-in (CI) free space reference distance model,  expressed as
\begin{equation}
\begin{split}
    \text{PL}^\text{CI}_{a} [\text{dB}]&=10n_{a}\log_{10}\frac{d}{d_0}+\text{FSPL}(d_0)+\chi_{a}
\end{split}
\end{equation}
where $a$ could either be “best” or “omni”. Besides, $n_{a}$ stands for the path loss exponent (PLE) of the CI model. Moreover, $d$ is the Euclidean distance between Tx, and Rx and $d_0$ denotes the reference distance, which is set as \SI{1}{m} in this work. Moreover, $\chi_a$ is the shadow fading term, which is modeled as a Gaussian distributed random variable. Furthermore, $\text{FSPL}(d_0)$ is the free space path loss at the reference distance $d_0$, which is calculated by using the Friis' law
\begin{equation}
    \text{FSPL}(d_0,f)=-20\times\log_{10}\frac{c}{4\pi fd_0}
\end{equation}
where $f$ denotes the frequency, and $c$ is the velocity of light.
\par The path loss and shadow fading results are calculated and summarized in Table~\ref{tab:char}, based on which we can elaborate several observations as follows. First, the best direction path loss values are larger than the omnidirectional ones, since the best direction path loss only involves influences of partial MPCs. Second, the PLE of best direction path loss is slight larger than 2, which is the PLE of FSPL, due to misalignment of Rx antennas. In contrast, the PLE of omnidirectional path loss is very close to 2, indicating that this scenario is very dominant by the LoS path. Third, weak shadow fading effects are observed in the atrium scenario, with standard deviation of the shadow fading as \SIrange{1}{2}{dB}.
\subsection{K-factor}
\par The K-factor is calculated as the ratio between power of the strongest cluster and power of other clusters. The measured K-factor values are fitted with log-normal distribution and the results are shown in Table~\ref{tab:char}. First, since only 1 cluster is observed at some Rx locations, such Rx 1, 2, etc., the K-factor values at these points are not calculated. Second, it can be observed that serveral Rx points have lower K-factor values than others, namely Rx 7 and Rx 19-21. The reason behind this is that these Rx points are close to the corners of the atrium, receiving significant reflections and scattering from the metal window frames around them. Third, the average value of the K-factor we measured is \SI{12.82}{dB}, indicating strong dominance of the LoS path. 
\subsection{Delay and Angular Spreads}
\par Since multipath components travel with different propagation distance and directions, their power disperses in both temporal and spatial domains, which can be characterized by delay and angular spreads. The delay spreads (DS), azimuth spreads of arrival (ASA) and the elevation spreads of arrival (ESA) are summarized in Table~\ref{tab:char}, from which we make several observations. First, it can be observed that the delay and angular spreads generally increase as the separation distance between Tx and Rx increases. This is reasonable since as the Rx moves farther away from Tx, the power of the LoS path decreases and influences of other NLoS paths are more significant, resulting in larger delay and angular spreads. Second, the observed ESA values are very small, around 1-$7^\circ$, since there is no roof in this outdoor atrium environment and propagations of MPCs happen mostly in the azimuth plane. Third, the mean values of delay and angular spreads are \SI{12.58}{ns} for DS, $12.02^\circ$ for ASA, and $3.63^\circ$ for ESA, respectively. These values are smaller than typical values in frequency bands below \SI{100}{GHz}, as specified in standardization files by 3GPP~\cite{3gpp.38.901}, which reveals weaker multi-path effects in the THz band.
\subsection{Cluster Parameters}
\par Using the DBSCAN method described in Sec.~\ref{sec:mpcc}, the cluster parameters are calculated, including the number of clusters, cluster delay spread (CDS), cluster azimuth spread of arrival (CASA) and cluster elevation spread of arrival (CESA), as summarized in Table~\ref{tab:char}. As can be seen, the number of clusters is larger for Rx points in the middle, where Rx can receive significant scattering from metal window frames and pillars around it. Furthermore, compared to typical values in standardized indoor office scenario for frequency bands below \SI{100}{GHz}~\cite{3gpp.38.901}, less clusters are observed in the atrium scenario in the THz band, indicating strong sparsity in the THz band. Moreover, similar to delay and angular spreads, the intra-cluster delay and angular spreads are also smaller than typical values in lower frequency bands, revealing weaker dispersion of MPC power within clusters in the THz band.
\section{Conclusion}
\label{sec:conclude}
In this paper, we conducted a measurement campaign in the atrium of a building at \SIrange{306}{321}{GHz} using a VNA-based channel sounder. The measured data is processed through calibration, channel estimation and MPC cluster procedures, after which the parameters of MPCs and clustering results are obtained. Furthermore, we calculated and analyzed the channel characteristics, including the path loss, shadow fading, K-factor, delay and angular spreads, as well as cluster parameters. Specifically, the observations are summarized as follows.
\begin{itemize}
    \item The omnidirectional PLE of 2.01 and the mean K-factor value of 12.86 are measured, indicating the strong dominance of the LoS path.
    \item Compared to lower frequency bands, the conducted measurement campaigns produce much smaller delay and angular spreads. Specifically, delay spread is measured of \SI{12.58}{ns} on average, while azimuth and elevation angular spreads are measured to be $12.02^\circ$ and $3.63^\circ$, respectively. Therefore, the propagation is confined in both temporal and spatial domains.
    \item Very few clusters (less than 5) are observed in the measured outdoor atrium, revealing the strong sparsity of THz channels. 
\end{itemize}
\bibliographystyle{IEEEtran}
\bibliography{yuanbo}

\begin{thebibliography}{10}
\providecommand{\url}[1]{#1}
\csname url@samestyle\endcsname
\providecommand{\newblock}{\relax}
\providecommand{\bibinfo}[2]{#2}
\providecommand{\BIBentrySTDinterwordspacing}{\spaceskip=0pt\relax}
\providecommand{\BIBentryALTinterwordstretchfactor}{4}
\providecommand{\BIBentryALTinterwordspacing}{\spaceskip=\fontdimen2\font plus
\BIBentryALTinterwordstretchfactor\fontdimen3\font minus
  \fontdimen4\font\relax}
\providecommand{\BIBforeignlanguage}[2]{{%
\expandafter\ifx\csname l@#1\endcsname\relax
\typeout{** WARNING: IEEEtran.bst: No hyphenation pattern has been}%
\typeout{** loaded for the language `#1'. Using the pattern for}%
\typeout{** the default language instead.}%
\else
\language=\csname l@#1\endcsname
\fi
#2}}
\providecommand{\BIBdecl}{\relax}
\BIBdecl

\bibitem{chen2021terahertz}
Z.~Chen, C.~Han, Y.~Wu, L.~Li, C.~Huang, Z.~Zhang, G.~Wang, and W.~Tong,
  ``{Terahertz Wireless Communications for 2030 and Beyond: A Cutting-Edge
  Frontier},'' \emph{IEEE Communications Magazine}, vol.~59, no.~11, pp.
  66--72, 2021.

\bibitem{akyildiz2018combating}
I.~F. Akyildiz, C.~Han, and S.~Nie, ``Combating the distance problem in the
  millimeter wave and terahertz frequency bands,'' \emph{IEEE Communications
  Magazine}, vol.~56, no.~6, pp. 102--108, 2018.

\bibitem{rappaport2019wireless}
T.~S. Rappaport, Y.~Xing, O.~Kanhere, S.~Ju, A.~Madanayake, S.~Mandal,
  A.~Alkhateeb, and G.~C. Trichopoulos, ``{Wireless communications and
  applications above 100 GHz: Opportunities and challenges for 6G and
  beyond},'' \emph{IEEE access}, vol.~7, pp. 78\,729--78\,757, 2019.

\bibitem{priebe2011channel}
S.~Priebe, C.~Jastrow, M.~Jacob, T.~Kleine-Ostmann, T.~Schrader, and
  T.~K{\"u}rner, ``{Channel and propagation measurements at 300 GHz},''
  \emph{IEEE Transactions on Antennas and Propagation}, vol.~59, no.~5, pp.
  1688--1698, 2011.

\bibitem{priebe2013ultra}
S.~Priebe, M.~Kannicht, M.~Jacob, and T.~Kurner, ``{Ultra broadband indoor
  channel measurements and calibrated ray tracing propagation modeling at THz
  frequencies},'' \emph{Journal of Communications and Networks}, vol.~15,
  no.~6, pp. 547--558, Dec. 2013.

\bibitem{Eckhardt2021channel}
J.~M. Eckhardt, V.~Petrov, D.~Moltchanov, Y.~Koucheryavy, and T.~K{\"u}rner,
  ``{Channel Measurements and Modeling for Low-Terahertz Band Vehicular
  Communications},'' \emph{IEEE Journal on Selected Areas in Communications},
  vol.~39, no.~6, pp. 1590--1603, 2021.

\bibitem{ju2021millimeter}
S.~Ju, Y.~Xing, O.~Kanhere, and T.~S. Rappaport, ``Millimeter wave and
  sub-terahertz spatial statistical channel model for an indoor office
  building,'' \emph{IEEE Journal on Selected Areas in Communications}, vol.~39,
  no.~6, pp. 1561--1575, 2021.

\bibitem{ju2022sub}
------, ``Sub-terahertz channel measurements and characterization in a factory
  building,'' \emph{in Proc. of IEEE International Conference on Communications
  (ICC)}, Jun. 2022.

\bibitem{ju2021sub}
S.~Ju and T.~S. Rappaport, ``{Sub-Terahertz Spatial Statistical MIMO Channel
  Model for Urban Microcells at 142 GHz},'' \emph{in Proc. of IEEE Global
  Communications Conference (GLOBECOM)}, pp. 1--6, Dec. 2021.

\bibitem{abbasi2020double}
N.~A. Abbasi, A.~Hariharan, A.~M. Nair, A.~S. Almaiman, F.~B. Rottenberg, A.~E.
  Willner, and A.~F. Molisch, ``{Double directional channel measurements for
  THz communications in an urban environment},'' \emph{in Proc. of IEEE
  International Conference on Communications (ICC)}, pp. 1--6, Jun. 2020.

\bibitem{abbasi2021double}
N.~Abbasi, J.~G{\'o}mez, D.~Burghal, R.~Kondaveti, S.~Abu-Surra, G.~Xu,
  C.~Zhang, and A.~Molisch, ``{Double-Directional Channel Measurements for
  Urban THz Communications on a Linear Route},'' \emph{in Proc. of IEEE
  International Conference on Communications Workshops (ICC Workshops)}, pp.
  1--6, Jun. 2021.

\bibitem{chen2021channel}
Y.~Chen, Y.~Li, C.~Han, Z.~Yu, and G.~Wang, ``Channel measurement and
  ray-tracing-statistical hybrid modeling for low-terahertz indoor
  communications,'' \emph{IEEE Transactions on Wireless Communications},
  vol.~20, no.~12, pp. 8163--8176, 2021.

\bibitem{wang2022thz}
Y.~Wang, Y.~Li, Y.~Chen, Z.~Yu, and C.~Han, ``{0.3 THz Channel Measurement and
  Analysis in an L-shaped Indoor Hallway},'' \emph{in Proc. of IEEE
  International Conference on Communications (ICC)}, pp. 1--6, Jun. 2022.

\bibitem{li2022channel}
Y.~Li, Y.~Wang, Y.~Chen, Z.~Yu, and C.~Han, ``{Channel Measurement and Analysis
  in an Indoor Corridor Scenario at 300 GHz},'' \emph{in Proc. of IEEE
  International Conference on Communications (ICC)}, pp. 1--6, Jun. 2022.

\bibitem{yuanbo2022channel}
------, ``{Channel Measurement and Characterization with Modified SAGE
  Algorithm in an Indoor Corridor at 300 GHz},'' \emph{arXiv
  preprint:2203.16745}, 2022.

\bibitem{ester1996density}
M.~Ester, H.-P. Kriegel, J.~Sander, X.~Xu \emph{et~al.}, ``A density-based
  algorithm for discovering clusters in large spatial databases with noise.''
  in \emph{kdd}, vol.~96, no.~34, 1996, pp. 226--231.

\bibitem{3gpp.38.901}
3GPP, ``{Study on channel model for frequencies from 0.5 to 100 GHz},'' {3rd
  Generation Partnership Project (3GPP)}, Technical Report (TR) 38.901, {Dec.}
  2019, version 16.1.0.

\end{thebibliography}

\end{document}